\documentclass[twoside,twocolumn]{article}

\usepackage{lmodern}
\usepackage{microtype} % Slightly tweak font spacing for aesthetics

\usepackage[top=30mm,left=18mm,right=18mm,bottom=25mm,columnsep=18pt]{geometry} % Document margins
\usepackage[small,labelfont=bf,up,textfont=it,up]{caption} % Custom captions under/above floats in tables or figures
\usepackage{booktabs} % Horizontal rules in tables
\usepackage{float} % Required for tables and figures in the multi-column environment - they need to be placed in specific locations with the [H] (e.g. \begin{table}[H])
\usepackage{hyperref} % For hyperlinks in the PDF

\usepackage{balance}

\usepackage{paralist} % Used for the compactitem environment which makes bullet points with less space between them

\usepackage{abstract} % Allows abstract customization
\usepackage{titlesec} % Allows customization of titles
\renewcommand\thesection{\Roman{section}} % Roman numerals for the sections
\renewcommand\thesubsection{\Roman{subsection}} % Roman numerals for subsections
\titleformat{\section}[block]{\large\scshape\centering}{\thesection.}{1em}{} % Change the look of the section titles
\titleformat{\subsection}[block]{\large}{\thesubsection.}{1em}{} % Change the look of the section titles

\usepackage{fancyhdr} % Headers and footers
\usepackage[figuresright]{rotating}
\usepackage[british]{babel}
\usepackage{xcolor}
\usepackage{amssymb}

\usepackage[square,sort,comma,numbers,round]{natbib}
\newcommand{\ourmethod}{{\em PSF-Likelihood}}
\newcommand{\limamethod}{{\em Li\&Ma}}
\newcommand{\limafitmethod}{{\em Li\&Ma with fit background}}

\title{\vspace{-15mm}\fontsize{22pt}{10pt}\selectfont\textbf{\sf Extending the Li\&Ma method to include PSF information}} % Article title

\author{
\large
\textsc{M. Nievas-Rosillo and J.L. Contreras}\\[2mm] 
\normalsize Universidad Complutense de Madrid \\
\normalsize \href{mailto:mnievas@ucm.es}{mnievas@ucm.es} \quad
\normalsize \href{mailto:jlcontreras@fis.ucm.es}{jlcontreras@fis.ucm.es}
\vspace{-5mm}
}

\date{}

\begin{document}

\balance

%\maketitle % Insert title

\thispagestyle{fancy} % All pages have headers and footers

%----------------------------------------------------------------------------------------
%	ABSTRACT
%----------------------------------------------------------------------------------------

\twocolumn[
  \begin{@twocolumnfalse}
    \maketitle
\begin{abstract}
\noindent The so called Li\&Ma formula is still the most frequently used method for estimating the significance of observations carried out by Imaging Atmospheric Cherenkov Telescopes. In this work a straightforward extension of the method for point sources that profits from the good imaging capabilities of current instruments is proposed. It is based on a likelihood ratio under the assumption of a well-known PSF and a smooth background. Its performance is tested with Monte Carlo simulations based on real observations and its sensitivity is compared to standard methods which do not incorporate PSF information. The gain of significance that can be attributed to the inclusion of the PSF is around of 10\% and can be boosted if a background model is assumed or a finer binning is used. 
\end{abstract}

\hfill \textit{data analysis, statistics, likelihood, imaging observations, Li\&Ma}

\vspace{30pt}

  \end{@twocolumnfalse}
]

%----------------------------------------------------------------------------------------
%	ARTICLE CONTENTS
%----------------------------------------------------------------------------------------

%\begin{multicols}{2} % Two-column layout throughout the main article text

\section{Introduction}

%\lettrine[nindent=0em,lines=3]{T}{}
The statistical significance of an observation is a key issue in {\it signal starved} fields such as Imaging Atmospheric Cheren\-kov Telescopes (IACTs) Astronomy, and in general Very High Energy (VHE) Astronomy. It determines whether a given astronomical source has been detected or not, providing a probability for the excess being due to background fluctuations. It also limits how much detail can be recovered in spectra and light curves, because a minimum significance is usually required for each spectral or light curve point to be accepted. Finally, it also plays an important role when the goal is to set upper limits for non-detected sources. In this case, the sensitivity of the method determines how constraining the upper limit is.

Until the publication of the classical article by Li\&Ma \cite{LiMa1983}, several approaches to define the significance of astronomical observations had been used in VHE observations. As shown in that article, most of them were based on incorrect statistical hypotheses, and thus yielded unexpected widths of the significance distributions when they were tested with Monte Carlo (MC) simulations. In their article, Li\&Ma proposed a robust and reliable method for estimating that significance. Since at that time VHE instrumentation had very limited angular resolution, the method was designed as an event counting technique which makes very little use of the instrument resolution, given by its Point Spread Function (PSF), and background distribution. Therefore, the sensitivity achieved should be worse than the one of methods that incorporate that information.

The \limamethod\ method, which shall be known as just \limamethod\ in the rest of this work, is a particular case of a more general family of techniques based on maximum likelihood principles. Generalized maximum likelihood methods such as that implemented in \cite{FermiLikelihood} and \cite{Fermi2FGL} are sometimes difficult to implement. There have been general proposals such as \cite{KlepserLike1} to extend the \limamethod\ formula or include the effect of systematic errors (e.g. \citep{Dickinson} and \citep{Spengler}). Still, the use of general likelihood methods in IACT Astronomy is restricted in practice to special analyses such as sky maps \cite{KlepserLike2} or spectral line studies in Dark Matter searches, as seen in \cite{SeguePaper} and \cite{JelenaSegue}. Nevertheless, even if they risk losing robustness and stability, they are usually more sensitive than event counting methods.

In this article, a simple technique that takes into account the a priori knowledge of the instrumental PSF is presented and characterized in detail, under the  assumption of a smooth background for which dedicated measures are available. Although the method is applicable to a wide range of situations, it has been tested in our field of interest: VHE observations. It can be understood as a generalization of the \limamethod\ method or a particular application of that proposed in \cite{KlepserLike1} to a specially relevant case: the search for one isolated point source in the field of view (FoV), which is the common case in extragalactic observations with the current sensitivity of IACT experiments. A point source is defined as one whose angular size is smaller than the PSF of the instrument. Known as the \ourmethod\ method it can recover more information from the source of interest while keeping, at the same time, the simplicity of the standard \limamethod\ method. In order to check whether the statistical foundations of the technique are correct, and estimate its rejection power, it is tested with a set of {\it toy Monte Carlo} samples generated using real background and data from observations of the Crab Nebula performed by the MAGIC experiment (\cite{MAGICPerformance2011}). The comparison allows what can be gained from this kind of approach in a real situation to be gauged.

%------------------------------------------------

\subsection{Maximum likelihood with background estimation\label{section:maxlik}}

IACTs operate in harsh environments and their performance is highly dependent on the atmospheric and instrumental observing conditions. As a consequence the background affecting an observation is highly variable and is usually estimated jointly with the signal.
In the past, the observation time was divided  between ON observations, in which the telescope was pointed towards the source, and OFF observations, in which the telescope was pointed to an equivalent region with no  source present. Nowadays it is customary to use alternative methods that do not require dedicated OFF observations. This is the case of the {\it Wobble} method, in which the telescope is pointed to different positions at a small fixed distance from the source. The size of the IACTs Field of View (FoV) makes it possible  to take simultaneous ON and OFF data, as described in \cite{MAGICPerformance2011}, \cite{WobbleFomin} and \cite{WobbleTBretz2005}. Sometimes it is possible to define several OFF regions within the same field, but additional care should be taken to avoid counting events twice.

All the significance estimators tested in this article are based on a binned Maximum Likelihood Ratio approach, which tests an assumed null hypothesis against an alternative one, formulated as:

\begin{description}
  \item[\bf Null hypothesis, H0] ON and OFF regions contain no sources, only background.
  \item[\bf Alternative hypothesis, H1] While the OFF region only contains background, in the ON region there is, in addition, a source. 
\end{description}

A simple case, in which the result of an observation is a one-dimensional histogram showing the number of events detected as a function of the squared distance to the source, will be assumed. The number of events per bin will follow Poisson statistics, leading to the Likelihood function:

\begin{eqnarray}
\mathcal L (X|\Theta) = \prod_{i}^{N} \frac{f^{n_i}_i (\Theta) e^{-f_i (\Theta)}}{n_i!} 
\end{eqnarray}

where $\Theta$ is the parameter space for our model, $i$ is the bin index (for a total of $N$ bins), $n_i$ the number of events in bin $i$ and $f_i$ the value of the test model in the given bin.

It is often convenient to work with the negative logarithm of this function,

\begin{eqnarray}
\mathbf L (X|\Theta) &\equiv& - \log \mathcal L (X|\Theta) = \nonumber\\
 &=& - \sum_i^N n_i \log f_i (\Theta) - f_i (\Theta) - \log n_i!
\end{eqnarray}

Since the last term of the summation is only a normalization factor, which does not depend on the parameters of the likelihood ($\Theta$), it can be safely removed and the expression simplified to: 

\begin{eqnarray}
\mathbf L' (X|\Theta) = - \sum_i^N  n_i  \log f_i (\Theta) - f_i (\Theta)
\end{eqnarray}

Now the likelihood ratio $\lambda$ and its logarithm can be computed, giving:

\begin{eqnarray}
-2 \log \lambda &\equiv& -2 \log\left[\frac{\mathcal L_{H0}(X|\Theta)}{\mathcal L_{H1}(X|\Theta)}\right] \nonumber = \\
&=& 2\{\mathbf L'_{H0} (X|\Theta) - \mathbf L'_{H1} (X|\Theta)\}
\end{eqnarray}

From \cite{WilksLR} it is known that, when the null hypothesis is true, $-2 \log \lambda$ asymptotically follows a $\chi^2_r$ distribution for large event counts, where $r$ is the difference in the number of degrees of freedom between both hypotheses. This can be used to compute the probability of the observed excess being due to a background fluctuation. It can also be translated into a test statistics $TS = \chi^2_1$, where $\chi^2_1$ is the value of the  $\chi^2$ with one degree of freedom corresponding to the same probability as the original $\chi^2_r$. The accurate approximation proposed by Wallace \cite{WallaceApprox} can be used to compute the corresponding value in the limit of high $TS$, while its sign can be set from the sign of the event excess.

\subsubsection{The \limamethod\ method}

In the \limamethod\ method, where $r=1$, only one bin is defined in each, ON and OFF regions. Then $TS$ has an analytical expression, which is normally known as the Li\&Ma formula (see \cite{LiMa1983}, formula 17). It depends on $n_{on}$ and $n_{off}$, the number of ON and OFF events respectively and $\alpha$,  the ratio between the effective ON and OFF observation times. A source region must be selected {\it a priori} to count ON and OFF events, which must be done carefully to avoid losing sensitivity. It is usually chosen as the one giving the maximum significance in a test sample (typically a Crab Nebula test sample) taking into account the PSF of the instrument and the expected background statistics. 

\subsubsection{The \limafitmethod\ method}

The number of OFF events can also be obtained by incorporating information from a region larger than that considered in \limamethod, by fitting a background model against the data and integrating the model in the selected signal region. This method usually gives smaller statistical uncertainties, as it is in principle equivalent to having better OFF statistics. In order to use this model, one must be aware of any existing inhomogeneity in the camera or other gamma-ray sources which would introduce additional components in the background shape. An additional constrain exists if Wobble-mode observations are performed, as the wobble offset (distance between the source position and the actual pointing position) limits the maximum range of the fit that can be used without double-counting events.

We will call this variant hereafter the \limafitmethod\ method. It can be implemented by calculating modified $\alpha' \equiv \alpha \sqrt{\frac{(\delta n_{off})^2}{n_{off}}}$ and $n_{off}' \equiv n_{off}\frac{\alpha}{\alpha'}$ values, where $\delta n_{off}$ is the estimated $n_{off}$ uncertainty. In this case, $\delta n_{off}$ is no longer the Poisson based $\sqrt{n_{off}}$, but the total uncertainty  estimated using the fit covariance matrix. $\alpha$ is the actual ratio between the effective ON and OFF time. These new $\alpha'$ and $n_{off}'$ values can be inserted into the \limamethod\ formula to get the significance.

\subsubsection{Other background estimation methods}

There are other ways of increasing the effective statistics in the background region and thus to potentially improve the sensitivity. One clear example is to increase the number of OFF regions as discussed in section 2.3 of \citep{BergeFunkHinton_2006}. An example of the gain that can be obtained with this approach is shown in Table \ref{Table:LiMaOFFRegions}. The main advantage of this method is that all the positions remain symmetric with respect to the center of the camera and the relative radial response is the same as in the ON region, which means that the only assumptions that are needed are a radially symmetric camera response, no significant sky changes among the different OFF regions and no additional sources present in the selected OFF positions. These requirements are different from those required in \limafitmethod\ formula and the best solution would thus depend on the particularities of the given instrument. 

\begin{table}
\centering
\footnotesize
\begin{tabular}{r|*{6}{r}}
& \multicolumn{6}{c}{Number of OFF positions} \\
& 1 & 3 & 5 & 9 & 15 & $\infty$ \\
\hline
$S (\sigma)$ & $4.2$ & $5.2$ & $5.5$ & $5.8$ & $5.9$ & $6.1$ \\
& & $+24\%$ & $+31\%$ & $+36\%$ & $+39\%$ & $+44\%$ \\
\end{tabular}
\caption{Expected improvements from Li\&Ma Formula 17 with the number of OFF positions ($1/\alpha$) for the particular case $n_{OFF}=640/\alpha$ and $n_{ON}=160+640$.\label{Table:LiMaOFFRegions}}
\end{table}

Another example is the so called {\it Ring method}\citep{BergeFunkHinton_2006}. The main advantages of this method is that its symmetry properties make it less prone to systematic errors due to sky gradients. In principle it can be applied to any point of the FoV and it is conceptually similar to other aperture photometry methods widely used in Astronomy. The main drawback is that the response of the OFF region is no longer the same as in the ON region because each position inside the ring lies at a different distance from the center of the camera. It thus becomes necessary to model the camera response carefully, which complicates the evaluation of the observation significance. The comparison of \limafitmethod\ with this method, while possible, is out of the scope of this paper.

\subsubsection{The \ourmethod\ method}

The method proposed in this work has been known as the \ourmethod\ method. It considers not only the number of ON and OFF events, but also the differences between the shapes of the signal and background model, that is, how the ON and OFF events are distributed. Any existing excess produced by a source should follow the shape of the PSF to be considered as a signal. The statistical hypothesis can be rewritten as:

\begin{description}
\item[\bf Null hypothesis, H0]
The ON and OFF observations have the same origin, and therefore the same functional shape. Both can be explained with the same background model.
\item[\bf Alternative hypothesis, H1]
The ON and OFF samples have a different origin. OFF contains only background events, while ON also contains signal events. An additional PSF-like component is needed to explain the ON data.
\end{description}

In this method, both histograms are fitted at the same time with the models derived from the aforementioned hypotheses, using common parameters, taking into account the different observation times. The Likelihood in each hypothesis is then defined as the product of the Likelihoods for ON and OFF, obtaining a total $\mathbf L' (X|\Theta)$ as the sum of the $\mathbf L' (X|\Theta)$ for the ON histogram plus that for the OFF data. The probabilities are calculated from $-2\log \lambda$ and translated into significances. The approach is similar to that used in the \limamethod\ method, but the implementation is slightly different. 

\subsection{Analytic expressions in the limit of perfect background\\ \mbox{knowledge}}

Based on common principles, the \ourmethod\  method and the \limamethod's one converge to similar mathematical expressions at certain limits.  This is the case when the background is perfectly known (infinite statistics and precise modelling). In that limit the contributions to the likelihood ratio from the OFF region fit with the two models used in the \ourmethod\ approach cancel approximately, and it can be written as

\begin{eqnarray}
-2 \log \lambda = 2 \sum_i^N \left[ n_{i,on} \log \frac{f_{i,on}}{f_{i,b}} - (f_{i,on} - f_{i,b}) \right]
\end{eqnarray}

where $f_{i,on/b}$ denotes the value of the background+signal and background models evaluated in bin $i$ and $n_{i,on/off}$ the contents of  the bin in the real observations.

A similar expression can be derived from Li\&Ma formula 17 if we take $n_{off} = \hat{n_B}/\alpha$ and calculate the limit of $-2\log \lambda$ when $\alpha$ is very small (perfect background knowledge).

\begin{eqnarray}
\lim_{\alpha \to 0^+} [-2\log \lambda] = 2\left[n_{on} \log \frac{n_{on}}{n_{b}} - (n_{on} - n_{b}) \right]
\end{eqnarray}

While the formula are similar, the \ourmethod\ expression is more restrictive on what is called a {\it signal}. It does not simply  require differences between the ON and OFF histograms, but also that the excess behaves like the PSF of the instrument in each bin. In addition the \ourmethod\ method naturally incorporates the information contained in a wider region, removing also the need for a tight cut in the extension of the signal region which, once optimized, may significantly decrease the signal statistics.

As an alternative to using one single bin, the \limamethod\ could be applied to several bins of the $\theta^2$ histogram individually. In this case, the whole excess would be incorporated, but with a significant drawback. Since no particular PSF shape is assumed, $-2\log \lambda$ would asymptotically behave like a $\chi^2$ with $N$ degrees of freedom, where $N$ is the number of used bins. It would then lead to a low $TS$ once this $\chi^2_N$ is translated into $\chi^2_1$. This is not the situation in the \ourmethod\, where the use of a predefined PSF shape that can predict several bin contents does not increase the number of degrees of freedom.

\section{Method\label{Section:Method}}

The methods described above were compared for the case of the search for a point-like source using simulated ON and OFF $\theta^2$ samples. These samples, plotted as histograms, show the number of events recorded as a function of $\theta^2$, where $\theta$ is the angular distance to the assumed source position. An automatic pipeline worked on them taking two histograms as input, one used as a source template, the other as the background template, and a template for the PSF. Several samples are then simulated with different amounts of excess events (signal) and computed the significance obtained from each method.

\begin{figure}[ht] % htp
\centering
\includegraphics[width=1\columnwidth]{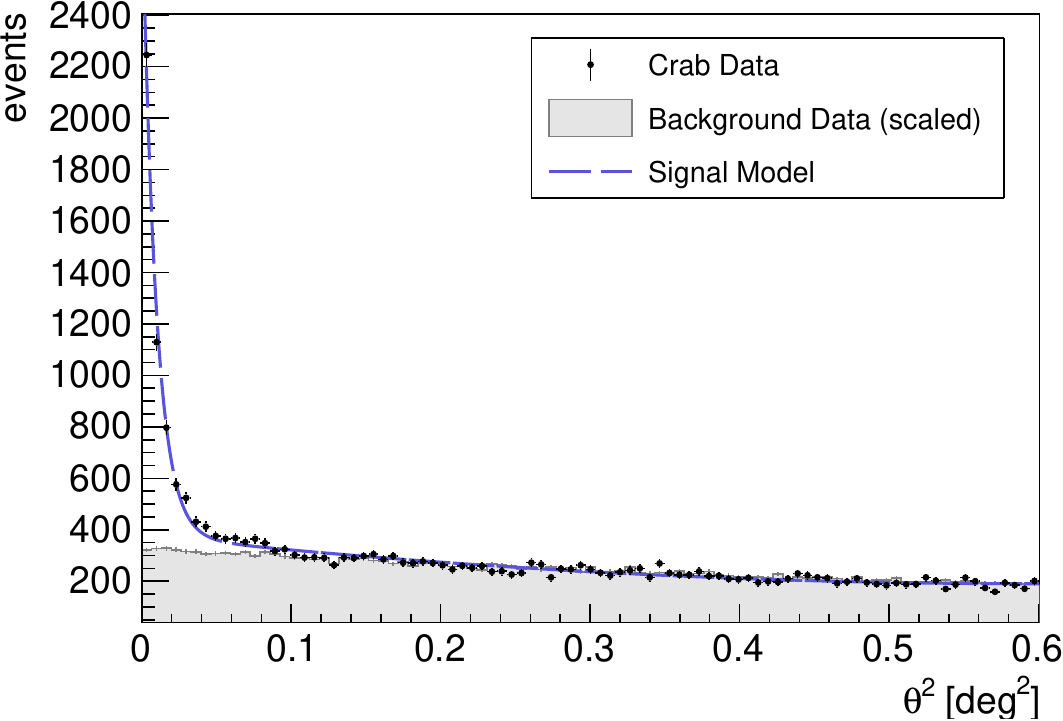}
\caption{Experimental $\theta^2$ distribution measured over a Crab Nebula sample with the MAGIC telescopes and scaled OFF data to be subtracted from the Crab Nebula data. The resulting distribution is used to generate the Monte Carlo simulated source samples on which we applied the methods.}
\label{fig:PureCrab}
\end{figure}

For the source template,a background subtracted signal from MAGIC observations of the Crab Nebula (the standard candle in VHE astronomy) were selected (see  Figure \ref{fig:PureCrab}).

\begin{figure}[ht] % htp
\centering
\includegraphics[width=1\columnwidth]{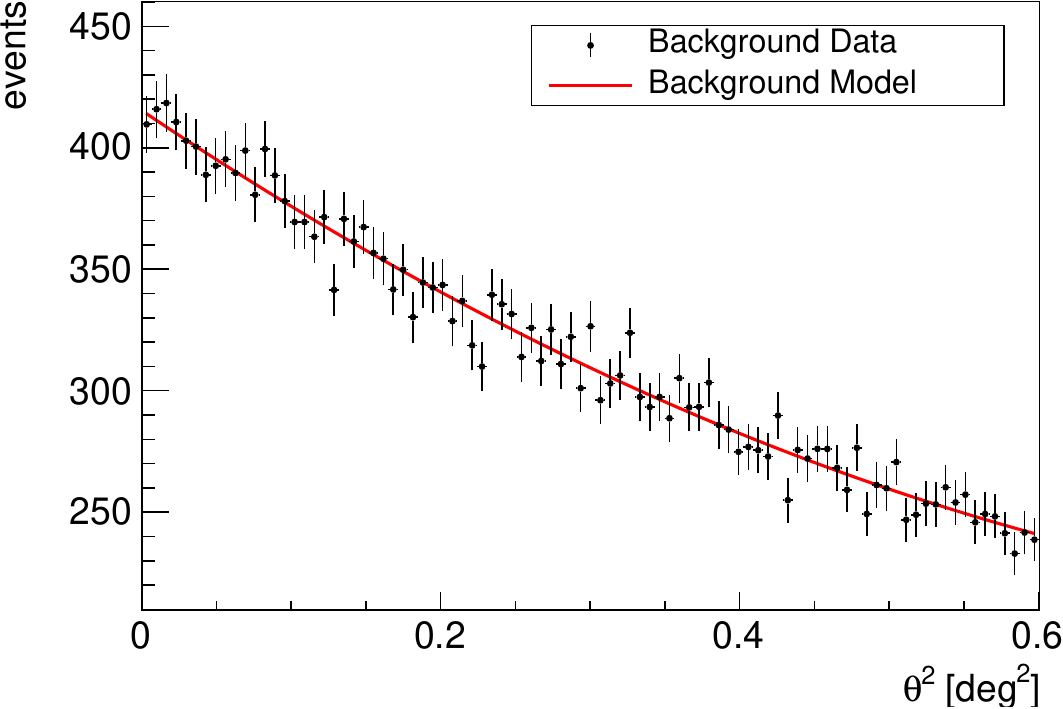}
\caption{Experimental $\theta^2$ distribution measured for a background region with the MAGIC telescopes using 3 simultaneous OFF regions. It was found that a 2nd order polynom was complex enough to reproduce the observed data, with a total $\chi^2 = 63.4$ for $88\ d.o.f.$. (In the $\theta^2 < 0.2$ range, using the same parameter values the result was $\chi^2 = 16.3$ for $28\ d.o.f.$).}
\label{fig:BackgroundFit}
\end{figure}

A different background sample was generated for each simulation, based on a background template with Poisson fluctuations in each bin. The number of simulated events was the same as in the real background scaled to the observation time. The template itself was obtained by fitting a second order polynomial to the histogram of a high statistics OFF sample. As seen in Figure \ref{fig:BackgroundFit} it reproduces the data used correctly. If, instead of a smooth template, a real background observation had been used directly as the model, it would have carried with it spurious fluctuations arising from the finite sample statistics, which the simulations would have propagated too, artificially increasing the total spread.

In order to build the PSF model two possibilities were explored, the case of the King function described in \cite{XMMCalibration} and a simpler Gaussian PSF. Both gave good results, with a minor improvement in the reproduction of the tails in the King function, at the cost of adding one more parameter to be optimized. Since the differences are rather small as regards the significance, the 1D Gaussian in $\theta$, with fixed width $\sigma$ was finally selected.

The values of $\sigma$ and the $\theta^2$ cut were optimized for the original Crab sample, so as to be in the best case scenario for all the methods. Using the isolated Crab signal and the background template, $3\cdot 10^6$ simulated ON and OFF samples were generated for $10$ different signal fractions (0\%, 0.2\%, 0.5\%, 1\%, 2\%, 3\%, 5\%, 8\%, 15\% and 50\%), covering a wide range of signal strengths. 

\begin{figure}[ht] % htp
\centering
\includegraphics[width=1\columnwidth]{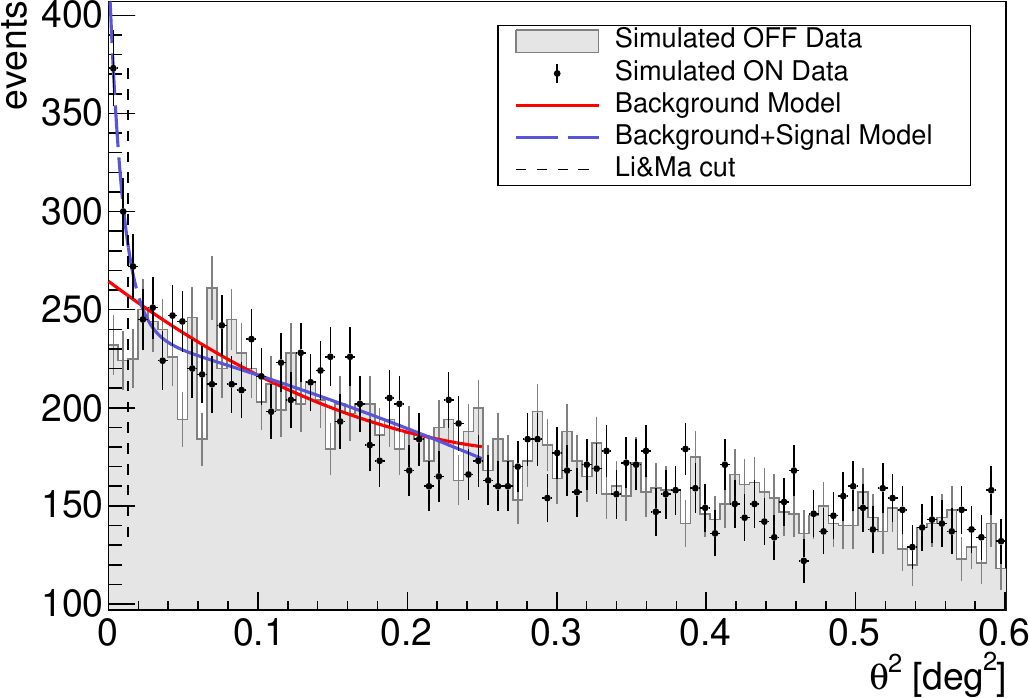}
\caption{Testing the method with simulated samples for an ON region with 8\% of excess events (black points and error bars) and a single OFF region (shaded region). The background-only model (red line) and background+signal model (blue line) are shown as reference. The dashed vertical line shows the region used by the \limamethod\ method to count events and calculate the significance. (For interpretation of the references to color in this figure legend, the reader is referred to the web version of this article.)}
\label{fig:SimulatedONOFF_80}
\end{figure}

In each simulation, the {\em H0} and {\em H1} statistical hypotheses were tested with the ON and OFF samples for the proposed methods and the significances were calculated. For the \ourmethod\ method this implied fitting the histograms to the models using a binned likelihood minimization with Poissonian errors, using the prescriptions from Section \ref{section:maxlik}. An example of the intermediate results can be seen in Figure \ref{fig:SimulatedONOFF_80}.

\section{Results\label{Section:Results}}

The main concern when testing a new detection technique is its statistical correctness. The distribution of significance provided by the method on pure background samples must follow the expected probability distribution. In our case, where the PSF model only adds one degree of freedom to those of the background model, the statistical distribution for background samples according to \cite{WilksLR} should be a $\chi^2_1$ just like in \limamethod.

\begin{figure}[ht] % htp
\centering
\includegraphics[width=1\columnwidth]{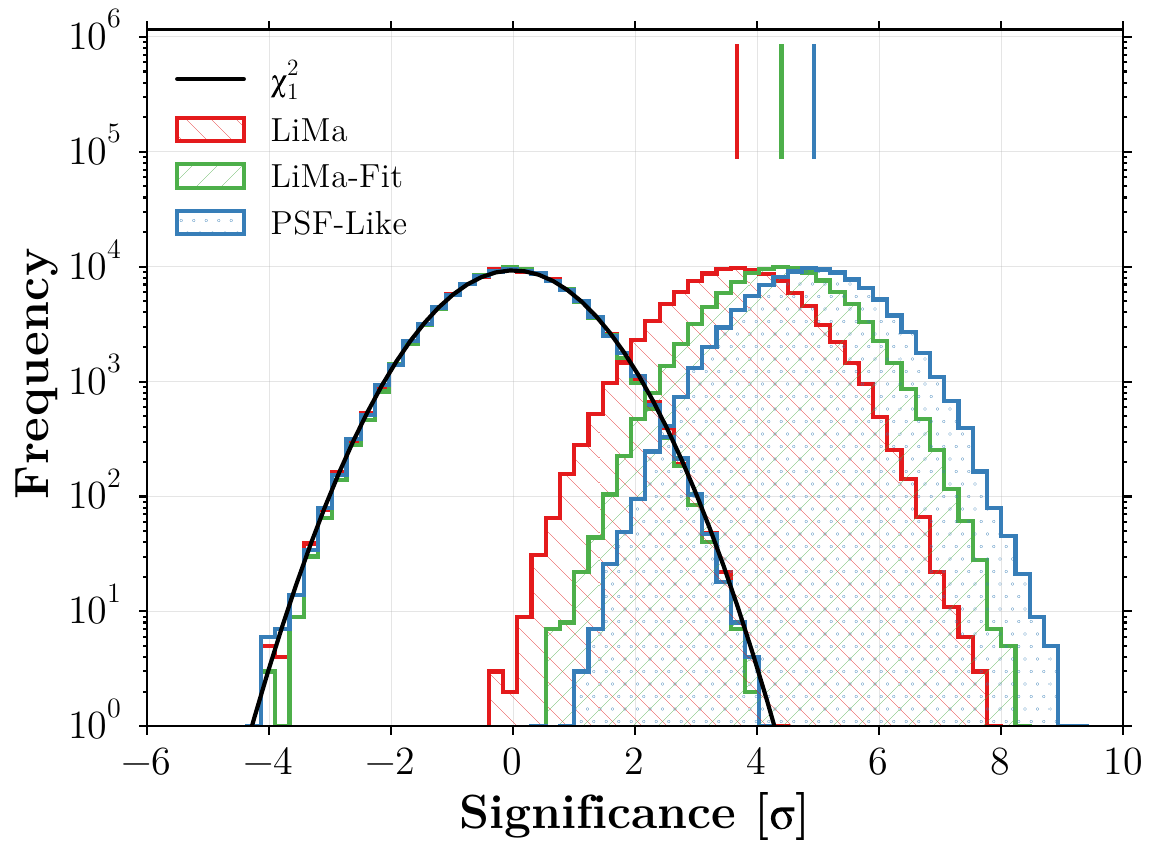}
\caption{Significance calculated on background-only Monte Carlo simulations and on a sample with 5\% signal with a single OFF region. The expected $\chi^2_1$ distribution for the background-only case is plotted as reference.}
\label{fig:StatTestBackground}
\end{figure}

From the left-hand curve of Figure \ref{fig:StatTestBackground}, it is seen that the three methods give statistically correct results when tested against background samples with random statistical fluctuations.

After validating the method statistically, it can be checked whether the method is competitive against existing ones. If this is the case, the mean significance provided by the method for samples containing signal should be higher. An example can be seen in the filled curves of Figure \ref{fig:StatTestBackground}, where the different methods are compared for a sample containing a 3\% of signal events. A drift towards higher significance values can be clearly seen for the proposed method. 

\begin{figure}[ht] % htp
\centering
\includegraphics[width=1\columnwidth]{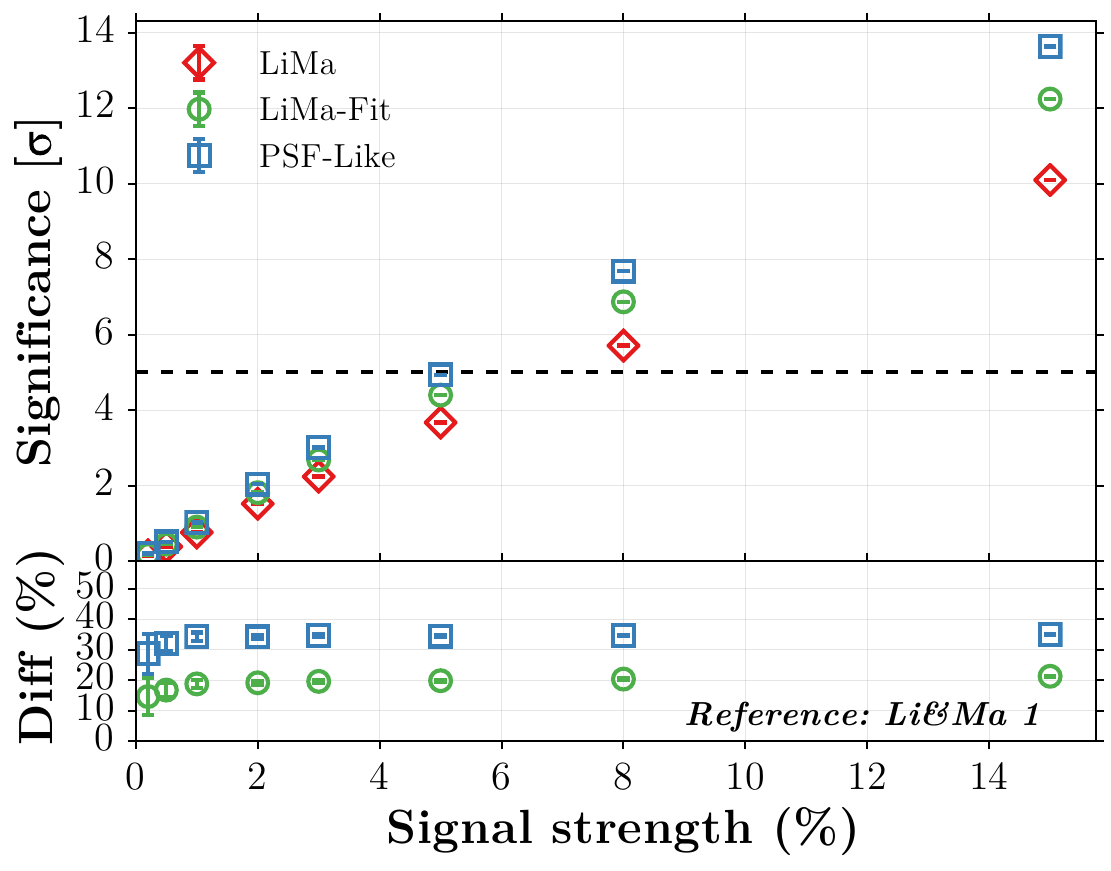}
\caption{Significance of different methods on Monte Carlo simulations. The $5\sigma$ limit (the usual detection threshold) is drawn as a horizontal dashed line. The differences ($\%$) always refer to {\limamethod}.}
\label{fig:StatTestSignal}
\end{figure}

A more complete comparison that covers all the different fractions of excess events simulated is presented in Figure \ref{fig:StatTestSignal}. From the figure, it seems evident that increasing the background statistics (using for instance \limafitmethod\ to extend the region used to calculate the background) helps to improve the sensitivity over the \limamethod\ method. In the same test, \ourmethod\ outperforms \limamethod\ and \limafitmethod\ for every step in the signal fraction, proving that taking into account the PSF also contributes to improving the sensitivity of the method.

It must be noted that the method allows, at the same time, the number of events detected from the source to be computed, which is simply related to  the normalization factor of the fit PSF, and its uncertainty. In that case, the improved sensitivity of the method is translated into smaller uncertainties for the number of excess events, and therefore the fluxes that can be computed from them.

This procedure could also be translated to data with finer energy bins, i.e. the source spectrum. The PSF to be used should then be optimized for each energy bin, as angular resolution usually depends strongly on energy.

In order to estimate how much the uncertainty in the fluxes can be reduced with this method, one may consider the extremely simplified case in which the background is perfectly known, so that $\sqrt{TS} \sim \frac{S}{\sigma(S)}$. Thus, a $35\%$ improvement in $\sqrt{TS}$ would translate into a $35\%$ decrease in the estimated uncertainty. Being more conservative and removing the part of the gain which comes from the improved background statistics, the remaining improvement would be of the order of $10\%$, as will be seen in section \ref{Section:RealBackground}.

\subsection{The method in the limit of low statistics\label{Section:LowStatistics}}

In \cite{LiMa1983}, Li \& Ma made an extensive study using MC simulations to detect the practical limits of the Likelihood Ratio approach, since the Wilks theorem \cite{WilksLR} only assures that the result is valid for high statistics. They found that the method is fairly robust, giving statistically accurate results with as few as 10 events in the ON and OFF samples. 

It should be kept in mind that \ourmethod\ has additional technical complications, which are not genuinely due to the method, but to the implementation. Instead of counting events, it tries to minimize a function, which is not always trivial and the algorithm might fail to converge due to local minima or wrong calculation of gradients, especially in the very low statistics regime. In order to check whether this technical problem could be a potential drawback of the method or not, MC simulations with the same parameters as before were carried out, using a reduced equivalent exposure, which would give rise to very low counts in the ON and OFF samples.

\begin{figure}[ht] % htp
\centering
\includegraphics[width=1\columnwidth]{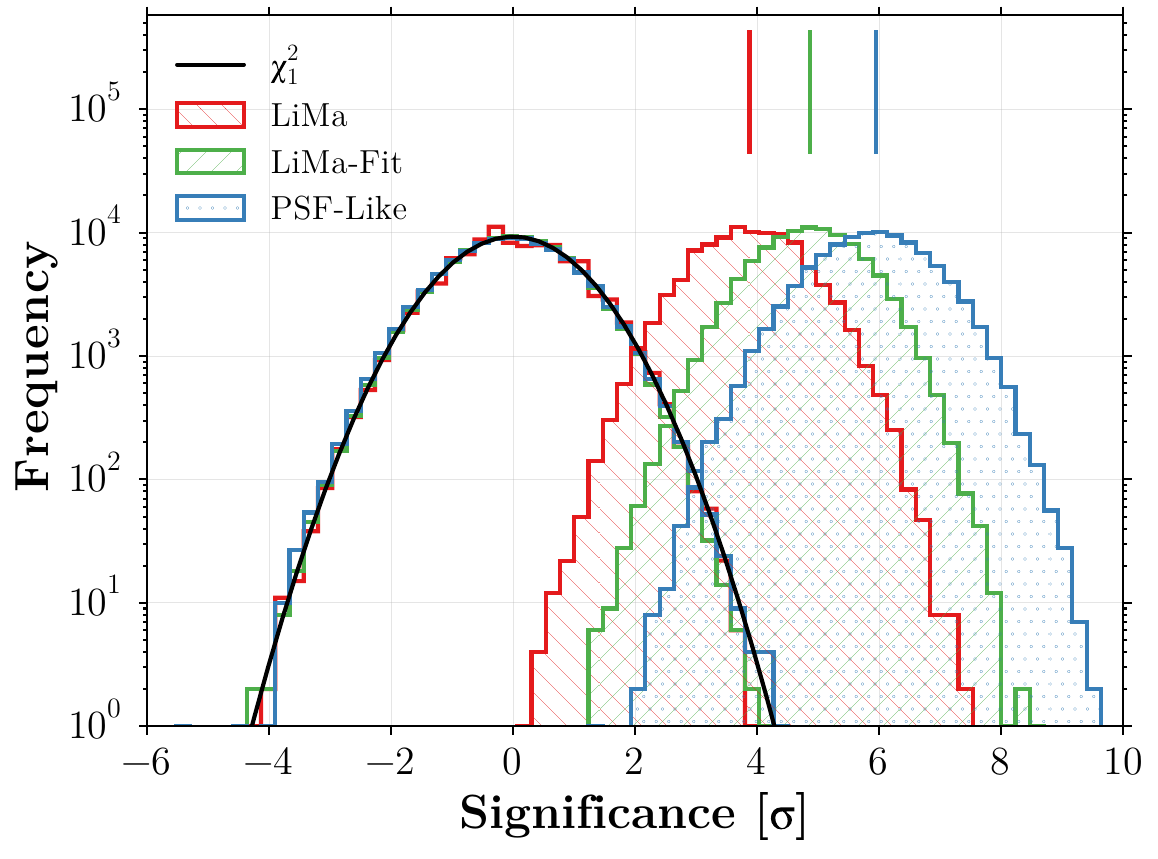}
\caption{Significance calculated on background-only Monte Carlo simulations and on a sample with 50\% signal with very low statistics per bin.}
\label{fig:StatTestBackground_lowstat}
\end{figure}

It was found that, even if the statistics are scarce (of the order of 10 events in the \limamethod-equivalent ON region), the method still works fairly well in most of the cases, with essentially no degradation in the estimated values and with only a minor fraction (of the order of $0.5\%$) of non-classifiable histograms. Figure \ref{fig:StatTestBackground_lowstat} shows the effect of event quantization around $TS=0$ in the \limamethod\ distribution.

\subsection{Non-optimal PSF model\label{Section:Systematics}}

An accurate knowledge of the PSF is not only important in \ourmethod\ but also in \limamethod\ when optimized cuts are used. In fact, in the latter, the estimation of the optimal cut to select the ON zone is usually done by evaluating both the PSF width and the number of background events. This should be done very carefully and blindly, otherwise it would bias the estimated significance.  For \ourmethod, a realistic PSF model is desirable to improve the sensitivity when a signal exists in the data, which has a significant impact on the discriminating power of the technique. Nevertheless the correctness of the PSF does not affect the statistical validity of the method, which depends mainly on the accuracy of the background model.

\begin{figure}[ht]
\centering
\includegraphics[width=1\columnwidth]{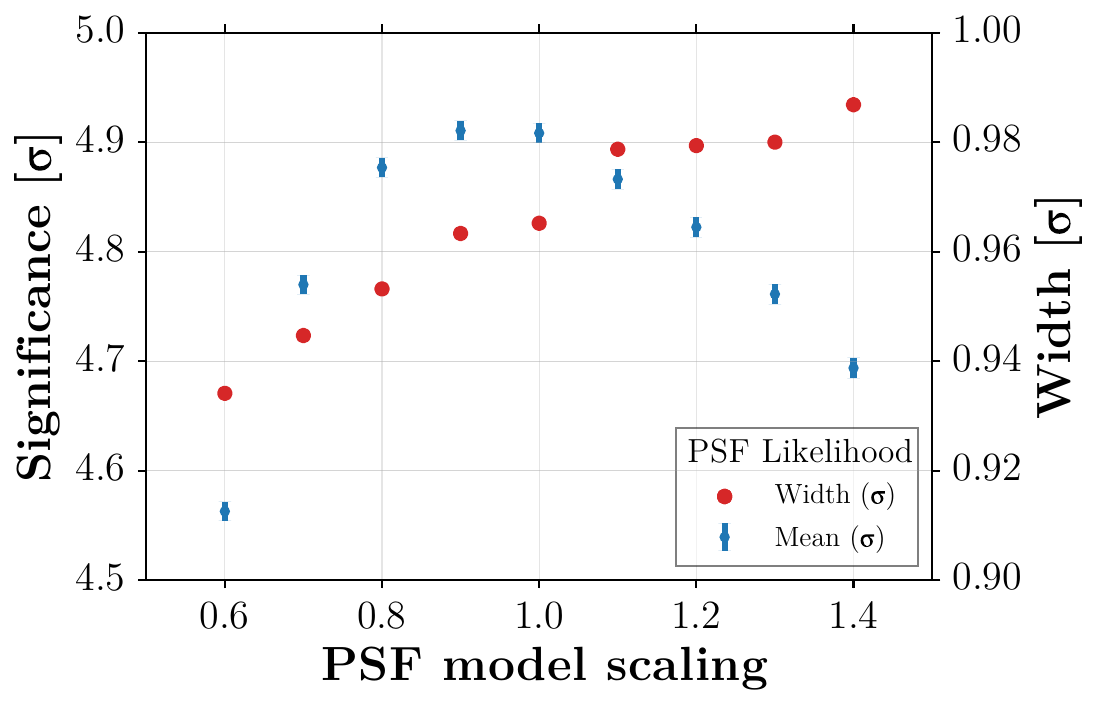}
\caption{Significance of PSF-Likelihood with a signal strength of $5\%$ with a systematically wrong PSF model.}
\label{fig:StatTestSignal_degradedpsf}
\end{figure}

In order to study the importance of a good PSF model, additional signal samples were generated and analyzed using a PSF template with a systematically wrong width (scaling from 0.6 to 1.4 times the nominal value). The resulting performance, shown in Figure \ref{fig:StatTestSignal_degradedpsf} and Table \ref{table:SubOptimalPSF}, can be compared with that from the standard analysis, represented by the case of a nominal PSF value. There is an obvious shift towards lower significance values of the distributions of $\sqrt{TS}$ with non-zero signal. Despite this, the degradation is never larger than $10\%$ even if the PSF width is wrong by a $40\%$.

\begin{table}
\centering
\begin{tabular}{r|*{5}{r}}
Signal & \multicolumn{5}{c}{PSF model scaling} \\
(\%) & 
\multicolumn{1}{r}{$0.7$} & 
\multicolumn{1}{r}{$0.9$} & 
\multicolumn{1}{r}{$1.0$} & 
\multicolumn{1}{r}{$1.1$} & 
\multicolumn{1}{r}{$1.3$} \\
\hline
$0.0$ & $0.0$ & $0.0$ & $0.0$ & $0.0$ & $0.0$ \\
$0.2$ & $0.2$ & $0.2$ & $0.2$ & $0.2$ & $0.2$ \\
$0.5$ & $0.5$ & $0.5$ & $0.5$ & $0.5$ & $0.5$ \\
$1.0$ & $1.0$ & $1.0$ & $1.0$ & $1.0$ & $1.0$ \\
$2.0$ & $2.0$ & $2.0$ & $2.0$ & $2.0$ & $2.0$ \\
$3.0$ & $2.9$ & $3.0$ & $3.0$ & $3.0$ & $2.9$ \\
$5.0$ & $4.8$ & $4.9$ & $4.9$ & $4.9$ & $4.8$ \\
$8.0$ & $7.4$ & $7.7$ & $7.7$ & $7.6$ & $7.4$ \\
$15.0$ & $13.2$ & $13.6$ & $13.5$ & $13.4$ & $13.1$ \\
$50.0$ & $36.8$ & $37.5$ & $37.3$ & $37.0$ & $36.0$ \\
\end{tabular}
\caption{Mean significance (in $\sigma$) when a suboptimal PSF shape is used. Errors are of the order of $\sim 0.0090$, with spread of less than $10\%$. \label{table:SubOptimalPSF}
} 
\end{table}

\subsection{Effects of binning\label{Section:Binning}}

Since the \limamethod\ formula uses one single bin in both the ON and OFF regions, the only discrimination power optimization that can be considered comes from the selection of their widths. It is limited by the statistics of events per bin when a tight bin is considered and the amount of background events when the bin is broader. On the other hand, in the \ourmethod\ method the ON region size is limited only by technical limitations of the instrument such as systematic uncertainties which may exist far from the center of the field of view. It seems therefore logical to think that the sensitivity of the method should improve as the bin width decreases due to a better description of the PSF. Although a detailed study of this effect would exceed the scope of this work, some checks were carried out.  

\begin{table}
\begin{center}
\begin{tabular}{r|r}
Bins & Significance ($\sigma$) \\
\hline
1 & $4.50 \pm 0.96$ \\
2 & $4.93 \pm 0.97$ \\
4 & $5.07 \pm 0.97$ \\
8 & $5.18 \pm 0.97$ \\
\end{tabular}
\end{center}
\caption{$\sqrt{TS}$ distribution mean and width for $5\%$ signal strength and different binning configurations for the \ourmethod\ method.}
\label{table:Binning_comparison}
\end{table}

Four different bin widths have been simulated and the resulting performance compared in Table \ref{table:Binning_comparison}, where it was found that the significance improves systematically with decreasing bin width. For a binning four times finer than the standard value (2 bins in $\theta<\theta^2_{cut}$), an additional improvement in \ourmethod\ of $\sim 5\%$ in $\sqrt{TS}$ would have been reached in all the simulations. It can be assumed that the performance could be improved even further with an unbinned likelihood approach, but its treatment is out of the scope of this paper as it would require knowing the precise position information for each event.

\subsection{Using real background data\label{Section:RealBackground}}

The study presented so far was centered in how to proceed when the background and signal behaviour can be described by smooth and simple models, with a few degrees of freedom. Figures \ref{fig:StatTestBackground}-\ref{fig:StatTestSignal_degradedpsf} are generated under this assumption. This is not always the case, and it can be argued that for some experiments the observed background cannot be easily predicted due to systematic effects and changing conditions in the instrument.

\begin{figure}[ht] % htp
\centering
\includegraphics[width=1\columnwidth]{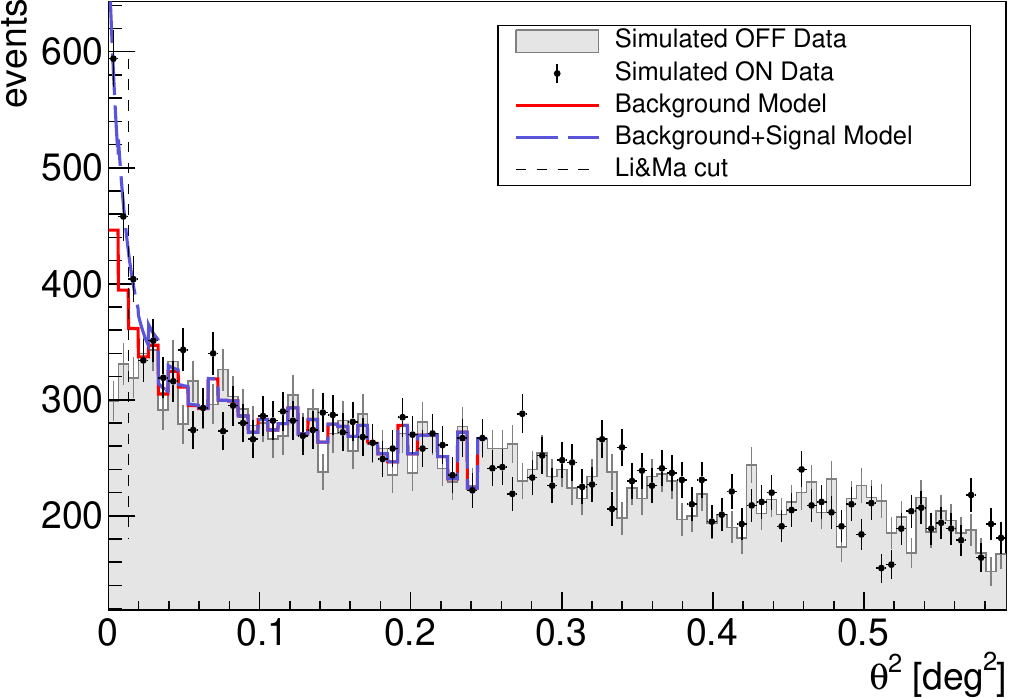}
\caption{Testing the method with simulated samples for an ON region with 15\% of excess events (black points and error bars) and a single OFF region (shaded region) using an analytic PSF and an empirical background shape (uncorrelated bin values).}
\label{fig:TestedMethod_realbackground}
\end{figure}

For these cases, the method can still be used with good performance. The idea is to replace the analytic function that provides the background shape (so far a polynomial) by a discrete function for which the value of each bin is totally independent, thus turning the bin values into uncorrelated variables. The Likelihood Ratio still behaves like a $\chi^2_1$ because the number of background parameters are the same in the null and alternative hypothesis and there is only one extra parameter to describe the signal. The sensitivity is slightly degraded as the background model is allowed to mimic the signal excess partially, as seen in Figure \ref{fig:TestedMethod_realbackground}.

\begin{figure}[ht] % htp
\centering
\includegraphics[width=1\columnwidth]{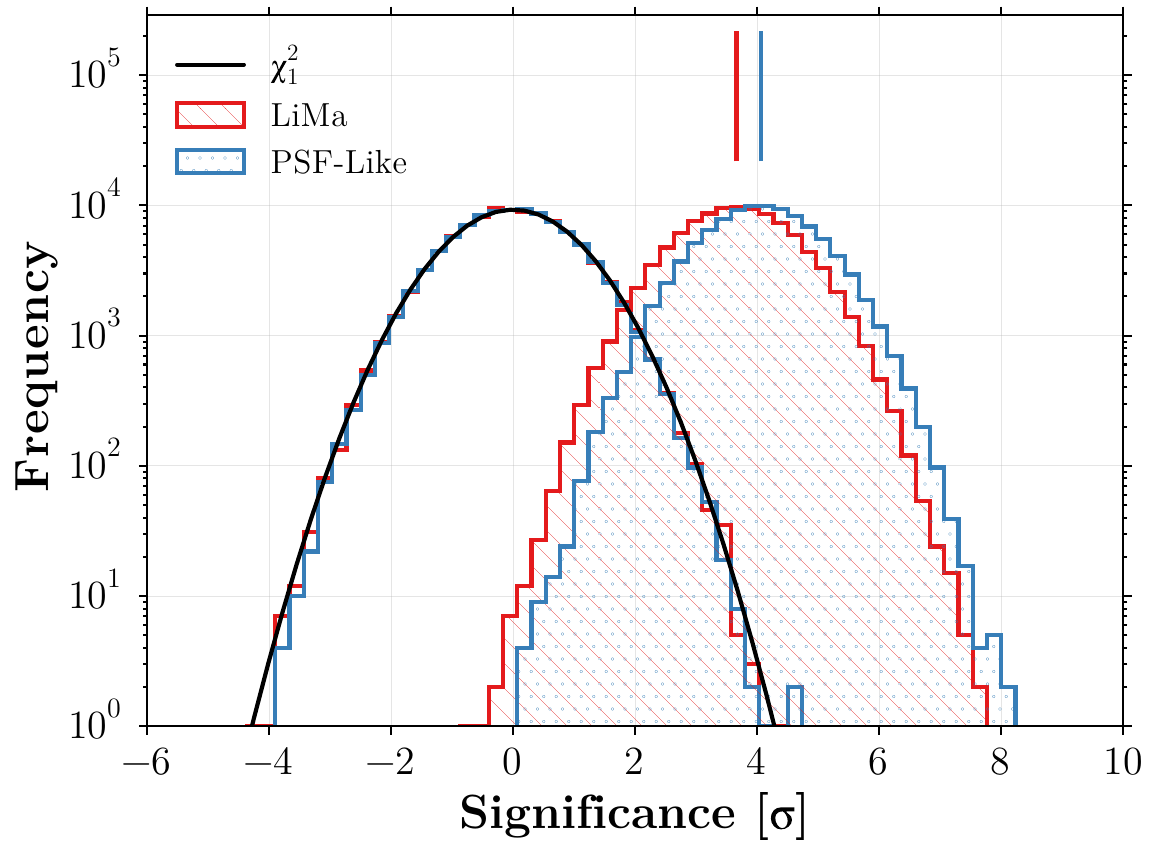}
\caption{Significance calculated on background-only Monte Carlo simulations and on a sample with 5\% signal.}
\label{fig:StatTestBackground_realbackground}
\end{figure}

\begin{figure}[ht] % htp
\centering
\includegraphics[width=1\columnwidth]{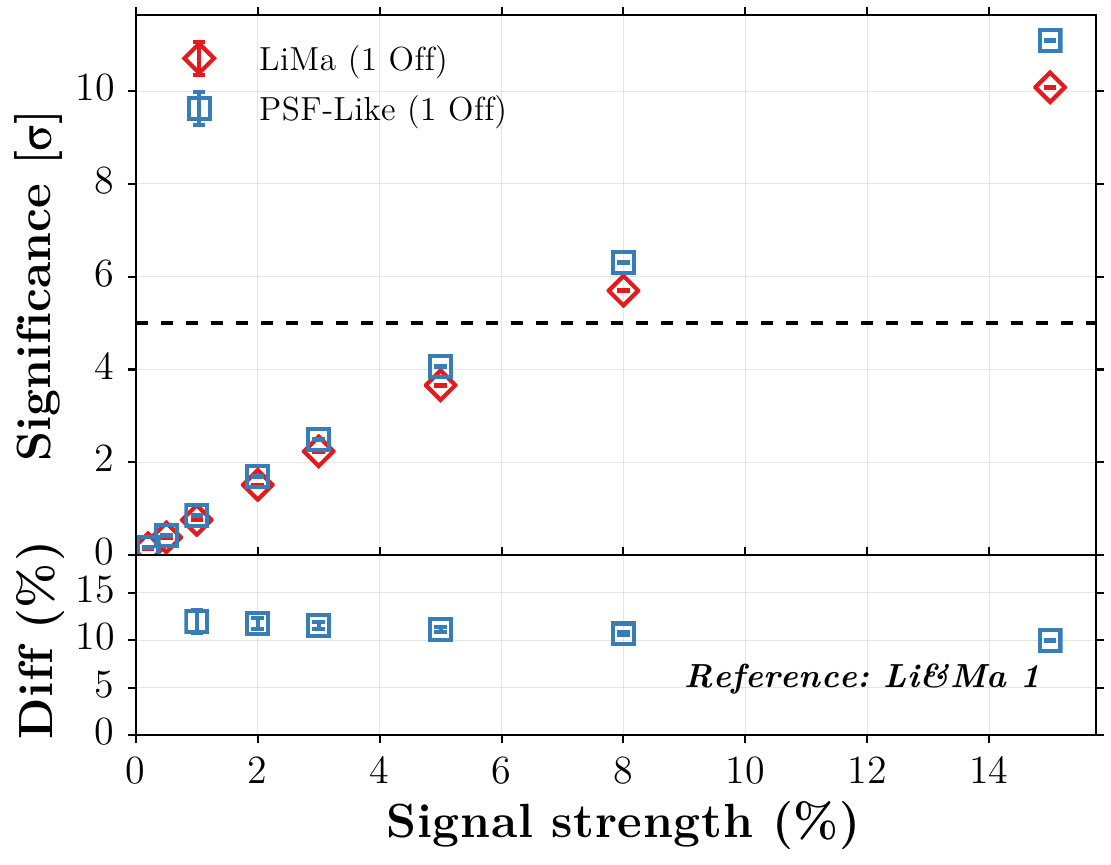}
\caption{Evolution of significance against signal strength and improvements on \limamethod\ with a single OFF region for the background model-less variant of \ourmethod.}
\label{fig:StatTestSignal_realbackground}
\end{figure}

Additional care should be taken during the implementation of this variant because the number of parameters for the background function is greatly increased, thus making the whole minimization  more complicated for standard algorithms such as {\tt Minuit} \cite{Minuit}. Once implemented, it can be seen that the method behaves correctly when tested against background only data (Figure \ref{fig:StatTestBackground_realbackground}) and still performs better than the standard Li\&Ma (Figure \ref{fig:StatTestSignal_realbackground}).

\section{Conclusions\label{Section:Conclusions}}

A possible implementation of the binned Likelihood Ratio method to estimate the significance of IACT observations of point-like sources, \ourmethod, has been described. The method considers measured $\theta^2$ distributions for an ON and OFF region and compares the likelihood that both of them are explained by the same background only model with that including also a source in the ON region. 
bv
When the method is tested on Monte Carlo simulations containing only background, it reproduces the expected $\chi^2_1$ significance distribution, proving that the chance probability of a false detection is correctly estimated.

If a certain amount of signal is included in the simulations, an improvement in sensitivity is found over other methods. Part of the gain can be attributed to the increased effective background statistics, but a significant fraction of it stems from the inclusion of the PSF in the method, as was shown in section \ref{Section:RealBackground}.

The method has been tested in different scenarios, comparing its sensitivity with the different techniques usually employed in the field. An alternative implementation, which does not need a careful modelling of the background, has been proposed and tested in section \ref{Section:RealBackground} for those cases in which the background behaviour is not predictable. At the cost of a more complex minimization process, it represents a possible very general worst case alternative implementation.

An additional test was carried out to check whether the performance of the method holds even if the PSF shape is not perfectly known or the reconstructed position of the events does not follow the expected $\theta^2$ distribution. It was found that even in this case, \ourmethod\ still outperforms \limamethod.

Finally it must be highlighted that the procedure proposed can be easily generalized to include additional information. As an example, two dimensional distributions of the events in the sky could be used incorporating the corresponding two dimensional PSF while maintaining the simplicity of the method. It is reasonable to assume, although it has not been tested, that this additional information would increase the discriminating power. The proposed method could also be used for any other imaging observation of point-like sources that incorporates an independent background observation. Also non-positional information such as the tagging variables usually employed in the IACT field to discriminate gamma rays from hadronic cosmic rays would be amenable to this kind of treatment.

\section*{Acknowledgments}

The authors would like to thank the MAGIC collaboration for the data samples included in the article. They would also like to thank Dr. A. Moralejo from {\it Institut de Fisica d'Altes Energies}, S. Bonnefoy, D. Carreto-Fidalgo and T. Hassan from {\it Universidad Complutense de Madrid} for the fruitful discussions held together. The authors also appreciate all the comments provided by MAGIC collaboration members, in particular the advice and comments from Drs J. Sitarek, O. Blanch, M. Doro and D. Paneque.

This work was supported by the Spanish MINECO under contract FPA2010-22056-C06-06 and MECD under FPU grant A-2013-12235. The analysis was performed using the open-source tools: ROOT Data Analysis Framework \cite{ROOTBrun}, Python and Matplotlib \cite{Hunter2007}.

\newpage
{\footnotesize
\bibliography{bibliography.bib}
}
%\end{multicols}

\end{document}